\documentclass[pdflatex,sn-mathphys]{sn-jnl}

\usepackage{xcolor}
\usepackage{graphicx}
\usepackage{hyperref}
\usepackage{ulem}
\usepackage{subfigure}
\usepackage{amsmath}
\usepackage{dsfont}
\usepackage{xcolor,colortbl}
\usepackage{diagbox}
\usepackage{amssymb}
\usepackage{bm}
\usepackage{accents}
\usepackage{float}

\theoremstyle{thmstyleone}

\theoremstyle{thmstyletwo}

\theoremstyle{thmstylethree}

\newcommand{\norm}[1]{\left\lVert#1\right\rVert}

\newcommand{\comment}[1]{}

\begin{document}

\title{CNN-based real-time 2D-3D deformable registration from a single X-ray projection}

\author*[1,2]{\fnm{F.} \sur{Lecomte}}\email{francois.lecomte@inria.fr}

\author[2,3]{\fnm{J.-L.} \sur{Dillenseger}}\email{jean-louis.dillenseger@univ-rennes1.fr}

\author[1,2]{\fnm{S.} \sur{Cotin}}\email{stephane.cotin@inria.fr}

\affil*[1]{\orgname{Inria}, \orgaddress{\street{1 Place de l'Hopital}, \postcode{67000}, \city{Strasbourg}, \country{France}}}

\affil[2]{\orgdiv{ICube}, \orgname{CNRS}, \orgaddress{\street{1 Place de l'Hopital}, \postcode{67000}, \city{Strasbourg}, \country{France}}}

\affil[3]{\orgname{Rennes University}, \orgaddress{\street{3 rue du Clos Courtel}, \city{Rennes}, \country{France}}}

\abstract{\textbf{Purpose:} The purpose of this paper is to present a method for real-time 2D-3D non-rigid registration using a single fluoroscopic image. Such a method can find applications in surgery, interventional radiology and radiotherapy. By estimating a three-dimensional displacement field from a 2D X-ray image, anatomical structures segmented in the preoperative scan can be projected onto the 2D image, thus providing a mixed reality view.
\textbf{Methods:} A dataset composed of displacement fields and 2D projections of the anatomy is generated from the preoperative scan. From this dataset, a neural network is trained to recover the unknown 3D displacement field from a single projection image.
\textbf{Results:} Our method is validated on lung 4D CT data 

at different stages of the lung deformation. 
The training is performed
on a 3D CT
using random (non domain-specific) diffeomorphic deformations, to which perturbations mimicking the pose uncertainty are added. The model achieves a mean TRE over a series of landmarks ranging from 2.3 to 5.5 mm depending on the amplitude of deformation.
\textbf{Conclusion:} In this paper, a CNN-based method for real-time 2D-3D non-rigid registration is presented. This method is able to cope with pose estimation uncertainties, making it applicable to actual clinical scenarios, such as lung surgery, where the C-arm pose is planned before the intervention.}


\maketitle

\section{Introduction}
\label{sec:introduction}
Laparoscopic surgery, interventional radiology and radiotherapy are among the most successful options for cancer therapy. These image-guided interventions are often totally or partially performed under fluoroscopic guidance, as this is for instance the case for lung surgery. On such moving and soft organs, the complexity of the intervention is obviously increased since the tumor position from the preoperative image becomes unknown or largely uncertain at the time of the procedure. The ability to manage organ motion (sometimes in real-time) is essential in the treatment outcome \cite{sharma_systematic_2022}. Therefore, estimating the new tumor position, in 3D and in real-time, from an intraoperative fluoroscopic image, can be a key improvement of these therapies. 
This is however very challenging as the combination of non-rigid deformation and reduced dimension of the intra-operative image make this problem ill-posed. 

Several surveys cover the associated challenges and possible approaches to such 2D-3D registration problems \cite{unberath_impact_2021,sotiras_deformable_2010}, as they can be handled in many different manners, given the application, imaging modality, parameter space, or optimization process. These registration methods can be divided in two groups: rigid and non-rigid. Rigid methods only compute a translation and a rotation while non-rigid methods are designed to estimate more complex displacement fields. Rigid registration methods are often needed to compensate for the unknown pose of the C-arm (and therefore the unknown projection from the CT to the fluoroscopic image) and can sometimes be sufficient when small deformations take place.  

In image-guided radiotherapy techniques, a study by Kilburn \textit{et al.} \cite{kilburn_image_2016} showed that correcting for the intra-operative deformation increased the 2-year survival rate from 64\
However, this procedure is invasive, increases the risks of complications to the patient and procedure time.

In the context of image-guided surgery, a study by Suzuki \textit{et al.} \cite{suzuki_video_1999} on Video-Assisted Thoracoscopic Surgery (VATs) found that failing to localize the tumor resulted in surgical conversion to open thoracotomy in 46\

The Bayesian approach presented in \cite{shieh_bayesian_2017} tackles the problem of 3D markerless tumor localization in radiotherapy. A pre-operative 4D CBCT is used to build a patient-specific respiratory model. During treatment, fluoroscopic images are acquired at frame rate of 5 images/s. A template matching algorithm is combined with a Kalman filter to predict the tumor position from the fluoroscopic image. Among the 13 cases in the study, the mean 3D error ranges from 1.6 to 2.9 mm. While these results are promising, the relatively high computation time and the need for a preoperative 4D CBCT  reduces the clinical usability of this method. 

The method in \cite{zhang_automatic_2020} uses a U-Net architecture combined with intensity-based registration and patient-specific biomechanical modeling to accurately localize liver tumors. The number of projections required, 20, requires specialized intraoperative imaging equipment, thus limiting the clinical applicability of the method.

Hirai \textit{et al.} developed a Neural Network based markerless tumor localization method from fluoroscopic images \cite{hirai_real-time_2019}. The network is trained on Digitally Reconstructed Radiographs (DRRs) generated from a 4D-CT augmented by small rigid transformations ($<2$ mm and $<1\deg$). The authors report a 3D tumor localization accuracy of $1.6$ mm, but the method requires the acquisition of a pair of fluoroscopic images.

The strategy proposed in \cite{foote_real-time_2019} consists in applying deformations to the preoperative CT data paired with DRRs to form a training and testing dataset for a deep learning algorithm. Displacement fields are obtained from a 4D-CT by selecting a rest CT phase and registering the other phases to the rest phase.

A PCA is computed from the displacements fields and sampled to generate the patient-specific deformation dataset. A neural network is then trained to estimate the PCA components 
from the DRR input.

The authors only evaluated their method on data from the PCA that was used to train the network.
No target registration error (TRE) is provided.

Most of the non-rigid registration methods described above assume a perfectly known intra-operative patient pose, which is needed to train the neural network. Many require a 4D-CT as preoperative image to generate patient-specific or problem-specific training data. Several methods also require multiple fluoroscopic images as input. In addition, existing methods are often limited in estimating a transformation from the 3D preoperative image space to the 2D fluoroscopic image. This is typically the case when a rigid transformation is estimated. However, this might prove insufficient when 3D tracking is required.

To address these limitations, we propose an approach that is robust against variation in poses, does not require a time series of preoperative images, and works with a single fluoroscopic image, in real-time. As a result our method is generic enough to address a variety of clinical applications where large, non-rigid deformations can occur.

To circumvent the need for a preoperative 4D CT, we implement a domain randomization solution to generate displacement fields using only the routinely acquired preoperative 3D CT. For this, we rely on the large deformation diffeomorphic metric mapping (LDDMM) framework (see section~\ref{sec:method-data-generation}) to enforce smoothness and invertibility of the displacement fields. 
We then introduce a small pose variation of the C-arm to account for the discrepancy between the planned and actual intra-operative pose. We generate DRRs from the deformed CT volumes. 

Our fully convolutional neural network then learns the mapping between the DRR and the 3D displacement field representing the rigid + non-rigid transform (see section~\ref{sec:method-network}).
The resulting 3D displacement field can be applied to the preoperative CT, or a segmentation extracted from it, to provide real-time tracking or visualization of internal structures (see section~\ref{sec:results}).

\section{Method}
\label{sec:method}
\begin{figure}[ht]
\centering
\includegraphics[width=\textwidth]{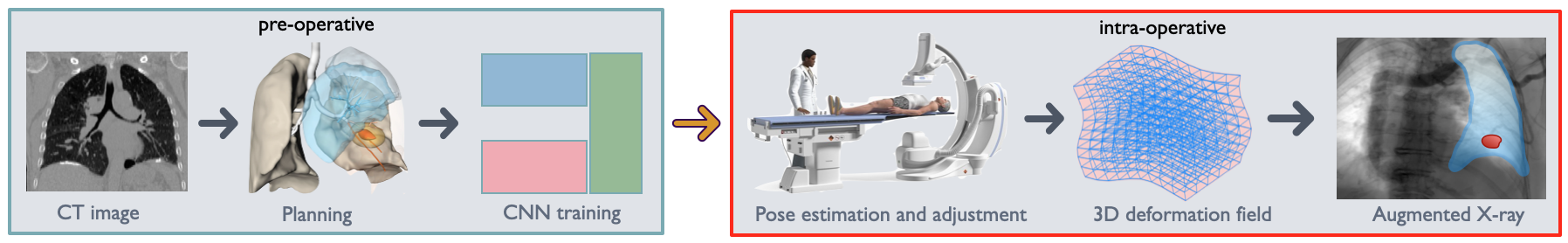}
\caption{Overview of our method: first, using a single 3D CT scan of the patient, we plan the intervention (i.e. the anatomical structures of interest are segmented and determine the optimal C-arm pose). Second, we train a neural network using randomized deformations and small randomized rigid transformations. Third, at the time of the intervention, the C-arm is adjusted to the planned pose, an X-ray image is acquired, and the network predicts in real-time the 3D deformation field from which we visually augment the fluoroscopic image.} \label{fig:pipeline}
\end{figure}

In order to be clinically relevant, our registration framework is based upon the most common steps of image-guided interventions (see figure~\ref{fig:pipeline}). At planning time, a 3D CT scan of the patient is acquired, structures of interest are segmented and the planning of the intervention is performed. At this stage, the pose of the C-arm and the pose of the patient are often chosen. Then, at treatment time, the patient and the C-arm are positioned following the treatment planning. To position the C-arm in the planned pose, a rigid registration must be performed. This step may be performed interactively \cite{Rouze2016} or automatically \cite{lee2022breathing}. In either case, the C-arm cannot be aligned perfectly to the targeted pose, resulting in a small pose error that must be accounted for by the registration method. Finally, a non-rigid registration method is often needed to compensate for the possible large non-rigid deformations and map the segmented structures of interest onto the fluoroscopic image. Remember that our objective is to recover a 3D displacement field, and not just a mapping from the 3D to the 2D image space. 

As discussed in section~\ref{sec:introduction}, non-rigid registration methods often assume no change between the C-arm pose at planning time and treatment time. However, because rigid registration methods have a limited accuracy, this assumption is invalid. For example, the mean 2D reprojection error of a state of the art automatic registration method (\cite{lee2022breathing}), is $7.8mm$. Consequently, our framework incorporates pose variation in the data generation process. At training time, the pose variation is composed with the displacement field to produce a target displacement field for the network to learn.

\subsection{Network architecture}
\label{sec:method-network}
Our network architecture is inspired by the work of Shen \textit{et al.} \cite{shen_patient-specific_2019} which goal is to reconstruct a CT volume from a fluoroscopic image. In this fully convolutional architecture, the direct translation from a dense 2D input to a dense 3D output is a key characteristic to learn fluoroscopy-to-CT mapping.  

In order to use this network for 2D-3D registration, the output of the network is now a 3D displacement field representing the total non-rigid + rigid transform of the preoperative CT to the intra-operative anatomy visible in the fluoroscopic image.

\begin{figure}[ht]
\includegraphics[width=\textwidth]{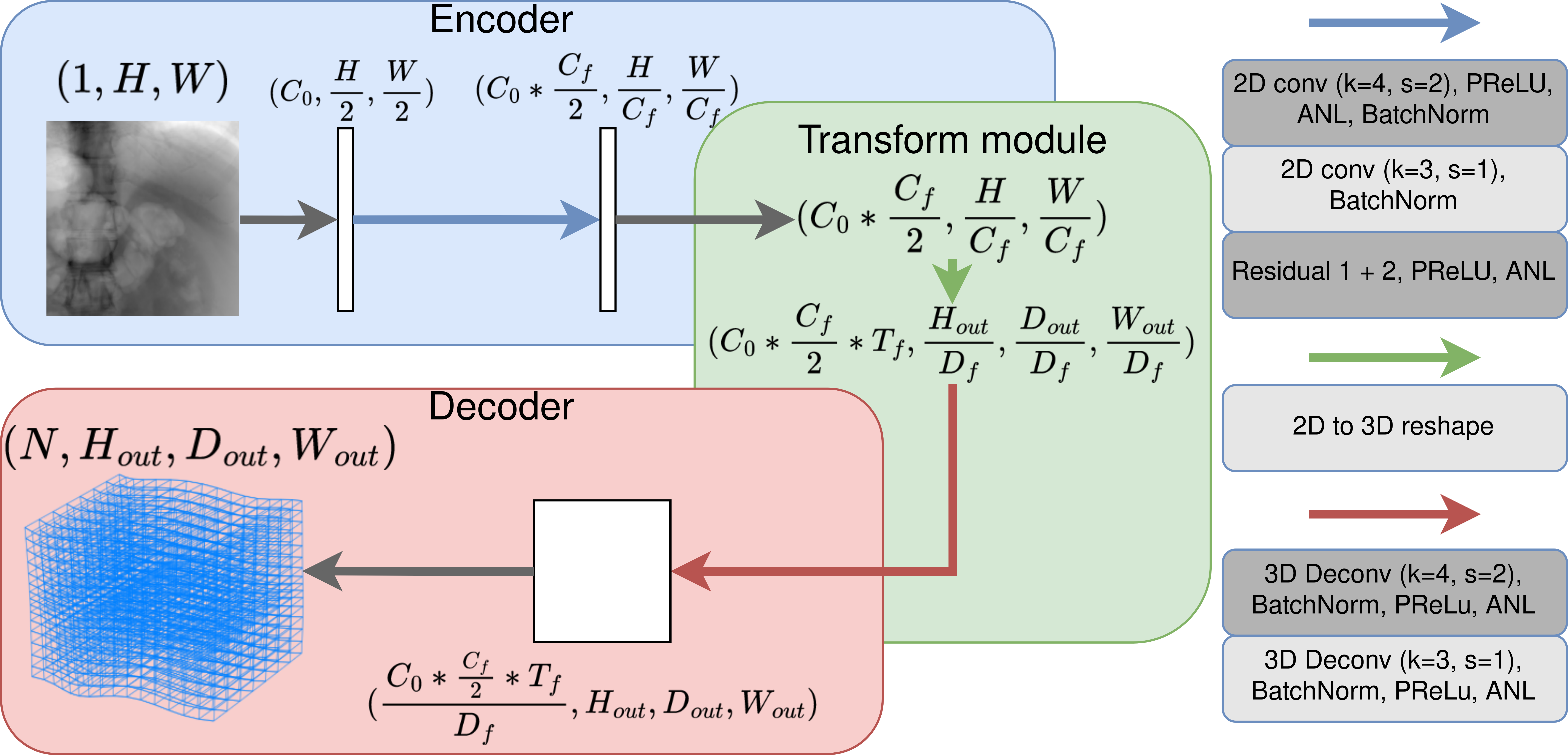}
\caption{Our network is based on an encoder-decoder architecture. The abbreviations $k$ and $s$ stand for kernel size and stride, respectively. The network takes as input one fluoroscopic image and outputs a sub-sampled 3D vector field on the CT volume. The central $(N_{c}-2)$ layers of the encoder are arranged in Residual Blocks. The data shape is fully described by: $C_f=2^{N_{c}/2}$, $D_f=2^{N_{dc}/2}$, and $T_f=(HWD_f^3)/(H_{out}D_{out}W_{out}C_f^2)$.\label{fig:network_architecture}}
\end{figure}

This allows the network to predict both global and local displacements in the preoperative CT from the intra-operative fluoroscopic image.
We also added several Adversarial Noise Layers (ANL) \cite{you_adversarial_2019} to regularize the latent space. The architecture of our network is summarized in figure~\ref{fig:network_architecture}.

The 2D input is an image of size $(1, H, W)$. The output 3D displacement field can also be considered as 3-channel volumetric image, of size $(3, H_{out}, D_{out}, W_{out})$. The first convolutional layer, represented by a gray arrow, with $k=4$ and $s=2$, transforms the input into a feature map with $C_0$ features. For every following Residual Block, H and W are divided by 2 and $C_0$ is multiplied by 2. The last convolutional layer, in gray, with $k=3$ and $s=1$, is not in a residual block and is followed by a BatchNorm, PRelu and ANL. The transform module performs a 2D to 3D reshape of the feature maps extracted by the encoder and passes them to the decoder. In the decoder, H, D and W are upscaled by a factor of 2 while the number of features is divided by 2 for every two of the $N_{dc}$ layers. An additional deconvolutional layer, in gray, with $k=1$ and $s=1$, transforms the $\frac{C_0*C_f}{D_f}$ output features into 2 channels. The 2 channels correspond to the two directions of displacement visible in the projection image. Predicting the displacement component perpendicular to the image plane is very challenging as a deformation in this direction leads to nearly no change in the fluoroscopic image. The network is supervised against the known 3D displacement field using an MSE loss, using AdamW \cite{loshchilov_decoupled_2018} ($\lambda=0.05$) as the optimizer. An ablation study (see table \ref{tab:ablation}) was performed to determine the best values for the  hyperparameters $N_{c}, N_{dc}, C_0, H, W, H_{out}, W_{out}, D_{out}$.

\subsection{Data generation}
\label{sec:method-data-generation}
To train our network, we generate a synthetic dataset composed of DRRs paired with 3D displacement fields, which are used to deform the preoperative CT $I_0$. From the deformed CT Scans, we generate DRRs using the DeepDRR algorithm \cite{unberath_deepdrr_2018}. We detail below these different steps and motivate our choices.

\paragraph{\textbf {Deformation generation:}}
To obtain smooth and invertible deformations $u(x)$, we generate Displacement Vector Fields (DVFs) following the LDDMM framework \cite{trouve_computing_2005}. Under this framework, the transformation $\phi$ that registers an image $I_0$ to an image $I_1$ is defined by $\phi(x)=x+u(x)$, of inverse $\phi^{-1}(x)=x-u(x)$, such that $\norm{I_0\circ\phi^{-1}(x)-I_1}^2$ is minimized. The vector field $u(x)$ is computed by integrating a velocity field $v(t,x)$ over time. A set of differential equations drive the evolution of $v(t,x)$ to satisfy the minimization condition.

In Durrleman \textit{et al.} \cite{durrleman_morphometry_2014}, the authors demonstrate that  $v(t,x)$ can be expressed as $v(t,x)=\sum_{k=1}^{N_{cp}}K(x,c_k(t))\alpha_k(t)$. In this formulation, $K$ is a element of a Reproducing Kernel Hilbert Space. From an implementation point of view, this permits to use a Gaussian kernel as $K$, and the solution of the minimization problem $u(x)$ is obtained in this case through the evolution of $\alpha_k$ and $c_k$.

\paragraph{\textbf{Domain randomization:}}
To implement Domain Randomization \cite{tobin_domain_2017} for deformation field generation, we randomly sample $\alpha_k(t)$ and $c_k(t)$ to generate a large variety of diffeomorphic displacement fields. This is done by first assigning a uniform probability distribution $U_{\alpha_k}(\mu_{\alpha_k}-\frac{w_{\alpha}}{2}, \mu_{\alpha_k}+\frac{w_{\alpha}}{2})$ for each $\alpha_k$ and $U_{c_k}(\mu_{c_k}-\frac{w_{c}}{2}, \mu_{c_k}+\frac{w_{c}}{2})$ for each $c_k$, with $(\mu_{\alpha_k}, \mu_{c_k})$ randomized for each sample. 
At each time step $t$, $\alpha_k(t)$ and $c_k(t)$ are sampled from $U_{\alpha_k}$ and $U_{c_k}$ to compute $v_t$. Because $v_t(x_t)$ is analytical, we can check that
$\norm{v_t(x_t)}_{W^{1,\inf}}<1$ where $\norm{v_t}_{W^{1,\inf,(\mathds{R}^N,\mathds{R}^N)}} = 
sup_{x\in \mathds{R}^N}(\left \lvert v_t(x_t) \right \rvert_{\mathds{R}^N}+
\left \lvert \nabla v_t(x_t) \right \rvert
_{\mathds{R}^N\times\mathds{R}^N})$ 
which ensures $u_t(x_t)=x_t+v_t(x_t)$ is a diffeomorphism \cite{Banyaga1997}
with $N$ the number of control points.  
We then obtain $x_{t+1}$ by $x_{t+1} = x_t+v_t$ before computing $v_{t+1}(x_{t+1})$. This way $u(x)$ is built iteratively from $t=0$ to $t=t_{max}$ and we can guarantee that $u$ remains a diffeomorphism. An additional check is performed at each time step to ensure that the volume of each voxel is not reduced below a given threshold $v_{thresh}$. With this process, we make sure that information is not lost when applying the displacement field to the image, and that the random DVFs remain diffeomorphisms. Through that process, we can apply the principles of domain randomization to our problem, and train the network in a generic way to ensure robustness and unbiasedness towards preferred directions of deformation. The generated DVFs are then applied to the preoperative CT via the \textit{grid\_sample} function in PyTorch.

\paragraph{\textbf {Digitally Reconstructed Radiographs:}} 
Because fluoroscopy involves ionizing radiations, obtaining a sufficient number of real images to train the network is not feasible. Instead, we generate Digitally Reconstructed Radiographs (DRR) using the DeepDRR framework \cite{unberath_deepdrr_2018}. This framework models the C-arm as a pinhole camera, parameterized by an intrinsic matrix $K$ and an extrinsic matrix
$
  E = \left[\begin{array}{ c | c }
    R & T \\
    \hline
    0 & 1
  \end{array}\right]
$
with $R$ a $(3,3)$ rotation matrix and $T$ a $(3,1)$ translation vector. Values for $T$ and $R$ are defined during the surgical planning stage (see figure \ref{fig:pipeline}). The intrinsic matrix is defined by detector characteristics, that are fixed for a given C-arm detector panel. The DRR projection $p$ is generated from $K$, $E$ and the CT volume $I_n$, as presented in \cite{unberath_deepdrr_2018}.

\paragraph{\textbf {Pose sampling:}}
While steps are taken to ensure the pose of the C-arm at treatment time is close to the planning pose, this step is still prone to errors (see \ref{fig:pipeline}), which is why the training data must include pose variations. A pose change matrix $P$ is parameterized in the same way as $E$, by a translation $T_P$ and a rotation $R_P$. The updated pose is given by $E'=PE$. 
$T_P$ if first sampled uniformly as a 2D vector parallel to the image plane with an amplitude between 0 and 1. $T_P$ is then scaled by $a\sim\mathcal{N_T}(0,\sqrt{T_{max}/2})$ with $T_{max}$ the amplitude of translations for the dataset. $R_P$ is sampled uniformly from the Haar distribution. $R_P$ is then converted to a rotation vector $r_P \in \mathds{R}^(3\times1)$. $r_P$ is then scaled in the same way as $T_P$, by $b\sim\mathcal{N_R}(0,\sqrt{R_{max}/2})$, with $R_{max}$ the amplitude of rotations for the dataset, and converted back to matrix representation. Finally, the tail of the normal distributions for $T_{max}$ and $R_{max}$ were cut for $\norm{T}>T_{max}$ and $\norm{R}>R_{max}$ to remove outliers.
$R_P$ and $T_P$ are sampled normally because the most likely pose of the C-arm at treatment time should be the planning pose, and poses far from the planning pose should be less frequent than poses close to the planning pose.

\paragraph{\textbf {Data generation summary:}}
A data sample $i$ is composed of $p_i$, $u_i(x)$, $P_i$. First, a displacement field $u_i(x)$ and a deformed CT $I_0'$ are generated via our domain randomization approach. Our pose sampling process is used to obtain $P_i$ and $E'$. Finally, knowing $I_0'$ and $E'$, we generate the projection $p_i$ via DeepDRR.

\section{Results \& discussion}
\label{sec:results}
\paragraph{\textbf {Experimental setup:}}
Using a lung 3D CT from the COVID-19-AR dataset \cite{Desai_chest_2020}, we generated 10,000 data samples. These samples were split between a training dataset containing 8,000 samples and a validation dataset containing 2,000 samples.

The data generation process described in \ref{sec:method-data-generation} was employed to generate the training/validation dataset. Assuming a maximum deformation amplitude of 30 mm, we then defined the associated Domain Randomization parameters. Thus, each of the 3 components of $\mu_{\alpha_k}$ and $\mu_{c_k}$ were drawn from uniform distributions $U(0,10)$ and $U(12,72)$. The values of $w_{\alpha}$  and $w_{c}$ were set to 0.001 mm and 0.6 mm respectively. Finally, the threshold $v_{thresh}$ was set to $0.75 mm^3$ and the displacement field was generated in $t_{max}=100$ iterations. For the DRR generation, the distance from the detector to the volume center was set to $\norm{T}=1500$ mm, and $R$ was defined such that the image plane normal was co-linear to the AP axis of the volume.
Finally, the parameters used for pose sampling were $T_{max}=17$ mm and $R_{max}=\frac{\pi}{4}$. To predict the total rigid + non-rigid transformation, the rigid transform was applied to a regular grid of points deformed by the non-rigid transform to obtain the final displacement field $u_{tot}$. This displacement field was used as the target displacement field for the network.

The network was trained for 75 epochs with an initial learning rate of $10^{-4}$ decreased to $10^{-5}$ after 30 epochs. Training took about 26 hours on a computer equipped with an AMD Ryzen 3950X CPU and an Nvidia Titan RTX GPU. The memory footprint of the network remains small, at 220 MB.

An ablation study was performed to determine the optimal network hyperparameters (see table \ref{tab:ablation}). Each hyperparameter was varied independently while the other parameters were fixed to their best values. The optimal hyperparameters are $N_{c}=6, C_0=64, H_{out}=64$. Some values combinations could not be tested due to memory constraints, eg. $H=512, C_{0}=64, H_{out}=64, N_{c}=6$.

\begin{table}
\centering
\parbox{.92\linewidth}{
\renewcommand{\arraystretch}{1.4}
\begin{tabular}{ |c|c| } 
\hline
\small Parameter considered & \small Validation loss \\ 
\hline
\small $(H,W)=(128,128)$; $(H_{out},D_{out},W_{out})=(32,16,32)$ & \small 49.22 \\
\hline
\small $(H,W)=(256,256)$; $(H_{out},D_{out},W_{out})=(64,32,64)$ & \small \textbf{38.41} \\
\hline
\small $N_{c}=6$ & \small \textbf{38.41} \\
\hline
\small $N_{c}=10$ & \small 53.69 \\
\hline
\end{tabular}
\vspace{3mm}
\caption{Ablation study reporting the lowest obtained validation loss for different values of $N_{c} = N_{dc}$, $C_0$, $H=W$, $H_{out} = W_{out}$, $D_{out} = \frac{H_{out}}{2}$.\label{tab:ablation}}
}
\hfill
\end{table}

\paragraph{\textbf {Results:}}
The accuracy of the network was evaluated on 6 anatomical landmarks covering both lungs. We measured the (3D) TRE between the predicted and ground truth positions but also the projection distance (PD) as suggested in \cite{van-de-Kraats_standardized_2005}.

On the validation dataset containing pose variations and domain randomized synthetic deformations, we obtained a mean TRE (mTRE) of 2.26$\pm$2.00 mm and a PD of 7.72$\pm$5.00 mm. In this validation dataset, landmark displacements range from 0.83 mm to 27.77 mm in 3D and 2.86 to 95 mm in 2D. This first result shows that our method is on par with the state of art even for large amplitudes of displacement while also accounting for pose errors.

To further validate our method, we tested it on a series of deformations from a 10-phase respiratory-correlated lung 4DCT provided by the 4D-Lung dataset \cite{hugo_longitudinal_2017}.
For the first testing dataset, the pose is not varied and the network is trained to recover only a non-rigid displacement.
For the second testing dataset, 20 poses were sampled for each of the 10 phases of the 4DCT to generate a testing dataset containing 200 samples.
The results for the first and second datasets are summarized in table \ref{tab:nr-results} and table \ref{tab:results}, respectively. 
We can see from the table \ref{tab:nr-results} that the domain randomization approach for displacement fields generation is justified, as the network performs very well (mTRE $\lt$ 3.16 mm) for realistic deformations even though it was trained on synthetic data. However, assuming that the pose is exactly known before the intervention is not realistic, hence the need to account for pose uncertainty.
The table \ref{tab:results} presents the results for a more challenging case, where the inputs are generated with a varying pose ($T_{max}=17mm$ and $R_{max}=\frac{\pi}{4}$). The network has thus been trained to recover such pose changes, and the accuracy remains high (mTRE $\lt$ 5.48 mm, mPD $\lt$ 2.31 mm) even though the 2D-3D registration task is now to estimate a rigid + non-rigid transform.

\begin{table}
\parbox{.40\linewidth}{
\renewcommand{\arraystretch}{1.25}
\begin{tabular}{ |c|c| } 
        \hline
        
        \footnotesize Phase \# & \footnotesize mean TRE (mm) \\
        \hline
        \small 0 & \small 0.8 $\pm$ 0.4  \\
        \hline
        \small 1 & \small 1.85 $\pm$ 1.28   \\
        \hline
        \small 2 & \small 3.02 $\pm$ 1.6   \\
        \hline
        \small 3 & \small 2.97 $\pm$ 2.57   \\
        \hline
        \small 4 & \small 2.96 $\pm$ 1.21   \\
        \hline
        \small 5 & \small 2.76 $\pm$ 1.67    \\
        \hline
        \small 6 & \small 3.16 $\pm$ 1.09    \\
        \hline
        \small 7 & \small 1.97 $\pm$ 0.84    \\
        \hline
        \small 8 & \small 1.84 $\pm$ 0.93    \\
        \hline
        \small 9 & \small 1.71 $\pm$ 0.53    \\
        \hline
\end{tabular}
\vspace{3mm}
\caption{Prediction results on 10 phases of a breathing cycle with only non-rigid displacements.\label{tab:nr-results}}
}
\hfill
\parbox{.55\linewidth}{
\centering
\renewcommand{\arraystretch}{1.25}
\begin{tabular}{ |c|c|c| } 
        \hline
        
        \footnotesize Phase \# & \footnotesize mean TRE (mm)  & \footnotesize mean PD (mm) \\
        \hline
        \small 0 & \small 4.48 $\pm$ 4.41  & \small 1.47 $\pm$ 0.88 \\
        \hline
        \small 1 & \small 3.44 $\pm$ 2.74  & \small 1.4 $\pm$ 0.85 \\
        \hline
        \small 2 & \small 4.85 $\pm$ 4.49 & \small 1.95 $\pm$ 1.27 \\
        \hline
        \small 3 & \small 5.12 $\pm$ 5.07 & \small 2.0 $\pm$ 1.42 \\
        \hline
        \small 4 & \small 5.03 $\pm$ 3.36 & \small 2.21 $\pm$ 1.34 \\
        \hline
        \small 5 & \small 4.29 $\pm$ 2.61 & \small 2.31$\pm$1.44  \\
        \hline
        \small 6 & \small 3.97 $\pm$ 3.07 & \small 1.87$\pm$1.12  \\
        \hline
        \small 7 & \small 4.16 $\pm$ 2.85  & \small 2.04$\pm$1.25  \\
        \hline
        \small 8 & \small 5.48 $\pm$ 3.88 & \small 2.1$\pm$1.64  \\
        \hline
        \small 9 & \small 3.66 $\pm$ 3.17 & \small 1.55$\pm$1.14  \\
        \hline
\end{tabular}
\vspace{3mm}
\caption{Prediction results on 10 phases of a breathing cycle when both rigid and non-rigid displacements are present. \label{tab:results}}
}
\end{table}

\begin{figure}
\centering
\includegraphics[width=\textwidth]{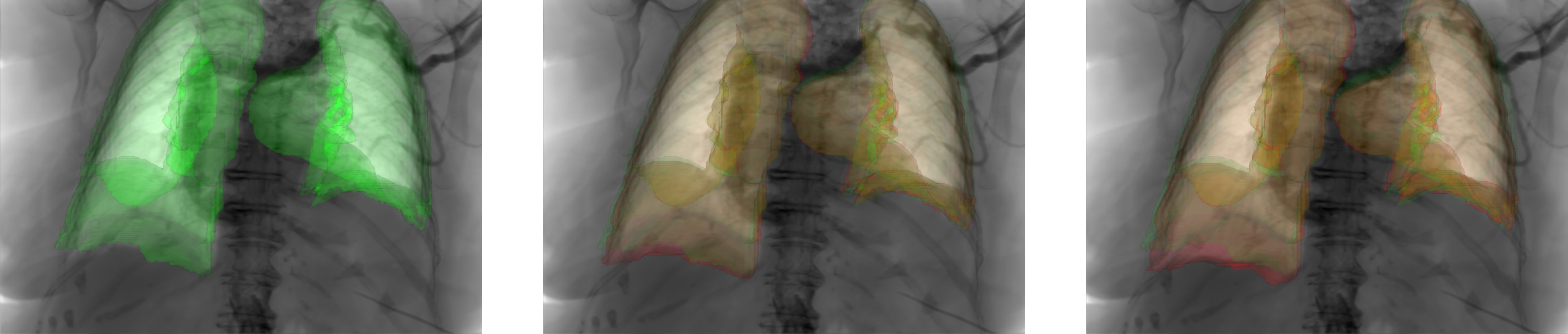}
\caption{Illustration of a lung segmentation deformed to match breathing motion, in red, superposed with the pre-operative segmentation, in green. The segmentations are overlayed on top of a DRR, simulating an augmented fluoroscopy.} 
\label{ fig:lung_AR}
\end{figure}

\section{Conclusion}
\label{sec:conclusion}
The objective of our work was to propose an accurate and real-time method able to recover a 3D displacement field from a single 2D fluoroscopic image.
We show that this ill-posed problem can be solved via modern deep learning techniques when associated with a comprehensive data generation pipeline. 

Our method only requires routinely acquired images (a single preoperative 3D CT and a single projection X-ray image at test time), and is robust to variations in pose, making it applicable to actual clinical scenarios. 
Our results show that the proposed method can estimate a 3D displacement field, even for structures deep into the tissues,
with an average accuracy of 4.45 mm. 

This level of accuracy is obtained at an update rate of about 20 Hz, 

sufficient for interactive visualization of an augmented fluoroscopic image as illustrated in figure  \ref{ fig:lung_AR}, or for real-time 3D navigation.

Our next steps will focus on the estimation of displacements that are perpendicular to the image plane, possibly through the use of a biomechanical model.

\backmatter

\bmhead{Acknowledgments}
The authors would like to thank Juan Verde, MD and Simon Rouze, MD for their valuable inputs during the development of this method. This work was funded by the French national research agency ANR (VATSOP project).

\bibliography{sn-article}


\begin{thebibliography}{25}
\ifx \bisbn   \undefined \def \bisbn  #1{ISBN #1}\fi
\ifx \binits  \undefined \def \binits#1{#1}\fi
\ifx \bauthor  \undefined \def \bauthor#1{#1}\fi
\ifx \batitle  \undefined \def \batitle#1{#1}\fi
\ifx \bjtitle  \undefined \def \bjtitle#1{#1}\fi
\ifx \bvolume  \undefined \def \bvolume#1{\textbf{#1}}\fi
\ifx \byear  \undefined \def \byear#1{#1}\fi
\ifx \bissue  \undefined \def \bissue#1{#1}\fi
\ifx \bfpage  \undefined \def \bfpage#1{#1}\fi
\ifx \blpage  \undefined \def \blpage #1{#1}\fi
\ifx \burl  \undefined \def \burl#1{\textsf{#1}}\fi
\ifx \doiurl  \undefined \def \doiurl#1{\url{https://doi.org/#1}}\fi
\ifx \betal  \undefined \def \betal{\textit{et al.}}\fi
\ifx \binstitute  \undefined \def \binstitute#1{#1}\fi
\ifx \binstitutionaled  \undefined \def \binstitutionaled#1{#1}\fi
\ifx \bctitle  \undefined \def \bctitle#1{#1}\fi
\ifx \beditor  \undefined \def \beditor#1{#1}\fi
\ifx \bpublisher  \undefined \def \bpublisher#1{#1}\fi
\ifx \bbtitle  \undefined \def \bbtitle#1{#1}\fi
\ifx \bedition  \undefined \def \bedition#1{#1}\fi
\ifx \bseriesno  \undefined \def \bseriesno#1{#1}\fi
\ifx \blocation  \undefined \def \blocation#1{#1}\fi
\ifx \bsertitle  \undefined \def \bsertitle#1{#1}\fi
\ifx \bsnm \undefined \def \bsnm#1{#1}\fi
\ifx \bsuffix \undefined \def \bsuffix#1{#1}\fi
\ifx \bparticle \undefined \def \bparticle#1{#1}\fi
\ifx \barticle \undefined \def \barticle#1{#1}\fi
\bibcommenthead
\ifx \bconfdate \undefined \def \bconfdate #1{#1}\fi
\ifx \botherref \undefined \def \botherref #1{#1}\fi
\ifx \url \undefined \def \url#1{\textsf{#1}}\fi
\ifx \bchapter \undefined \def \bchapter#1{#1}\fi
\ifx \bbook \undefined \def \bbook#1{#1}\fi
\ifx \bcomment \undefined \def \bcomment#1{#1}\fi
\ifx \oauthor \undefined \def \oauthor#1{#1}\fi
\ifx \citeauthoryear \undefined \def \citeauthoryear#1{#1}\fi
\ifx \endbibitem  \undefined \def \endbibitem {}\fi
\ifx \bconflocation  \undefined \def \bconflocation#1{#1}\fi
\ifx \arxivurl  \undefined \def \arxivurl#1{\textsf{#1}}\fi
\csname PreBibitemsHook\endcsname

\bibitem{sharma_systematic_2022}
\begin{barticle}
\bauthor{\bsnm{Sharma}, \binits{M.}},
\bauthor{\bsnm{Nano}, \binits{T.F.}},
\bauthor{\bsnm{Akkati}, \binits{M.}},
\bauthor{\bsnm{Milano}, \binits{M.T.}},
\bauthor{\bsnm{Morin}, \binits{O.}},
\bauthor{\bsnm{Feng}, \binits{M.}}:
\batitle{A systematic review and meta-analysis of liver tumor position
  variability during sbrt using various motion management and igrt strategies}.
\bjtitle{Radiotherapy and Oncology}
\bvolume{166},
\bfpage{195}--\blpage{202}
(\byear{2022})
\end{barticle}
\endbibitem

\bibitem{unberath_impact_2021}
\begin{botherref}
\oauthor{\bsnm{Unberath}, \binits{M.}},
\oauthor{\bsnm{Gao}, \binits{C.}},
\oauthor{\bsnm{Hu}, \binits{Y.}},
\oauthor{\bsnm{Judish}, \binits{M.}},
\oauthor{\bsnm{Taylor}, \binits{R.}},
\oauthor{\bsnm{Armand}, \binits{M.}},
\oauthor{\bsnm{Grupp}, \binits{R.}},
\oauthor{\bsnm{Kwok}, \binits{K.}},
\oauthor{\bsnm{Manfredi}, \binits{L.}},
\oauthor{\bsnm{Li}, \binits{C.}}:
The {Impact} of {Machine} {Learning} on {2D}/{3D} {Registration} for
  {Image}-{Guided} {Interventions}: {A} {Systematic} {Review} and
  {Perspective}.
Frontiers in Robotics and AI
\textbf{8}
(2021)
\end{botherref}
\endbibitem

\bibitem{sotiras_deformable_2010}
\begin{botherref}
\oauthor{\bsnm{Sotiras}, \binits{A.}},
\oauthor{\bsnm{Davatzikos}, \binits{C.}},
\oauthor{\bsnm{Paragios}, \binits{N.}}:
Deformable {Medical} {Image} {Registration}: {A} {Survey}.
IEEE Transactions on Medical Imaging
(2010)
\end{botherref}
\endbibitem

\bibitem{kilburn_image_2016}
\begin{barticle}
\bauthor{\bsnm{Kilburn}, \binits{J.M.}},
\bauthor{\bsnm{Soike}, \binits{M.H.}},
\bauthor{\bsnm{Lucas}, \binits{J.T.}},
\bauthor{\bsnm{Ayala-Peacock}, \binits{D.}},
\bauthor{\bsnm{Blackstock}, \binits{W.}},
\bauthor{\bsnm{Isom}, \binits{S.}},
\bauthor{\bsnm{Kearns}, \binits{W.T.}},
\bauthor{\bsnm{Hinson}, \binits{W.H.}},
\bauthor{\bsnm{Miller}, \binits{A.A.}},
\bauthor{\bsnm{Petty}, \binits{W.J.}},
\bauthor{\bsnm{Munley}, \binits{M.T.}},
\bauthor{\bsnm{Urbanic}, \binits{J.J.}}:
\batitle{Image guided radiation therapy may result in improved local control in
  locally advanced lung cancer patients}.
\bjtitle{Practical Radiation Oncology}
\bvolume{6}(\bissue{3}),
\bfpage{73}--\blpage{80}
(\byear{2016})
\end{barticle}
\endbibitem

\bibitem{adler_cyberknife_1997}
\begin{bchapter}
\bauthor{\bsnm{Adler}, \binits{J.R.}},
\bauthor{\bsnm{Chang}, \binits{S.D.}},
\bauthor{\bsnm{Murphy}, \binits{M.J.}},
\bauthor{\bsnm{Doty}, \binits{J.}},
\bauthor{\bsnm{Geis}, \binits{P.}},
\bauthor{\bsnm{Hancock}, \binits{S.L.}}:
\bctitle{The {Cyberknife}: {A} frameless robotic system for radiosurgery}.
In: \bbtitle{Stereotactic and {Functional} {Neurosurgery}},
vol. \bseriesno{69},
pp. \bfpage{124}--\blpage{128}
(\byear{1997})
\end{bchapter}
\endbibitem

\bibitem{seppenwoolde_treatment_2011}
\begin{botherref}
\oauthor{\bsnm{Seppenwoolde}, \binits{Y.}},
\oauthor{\bsnm{Wunderink}, \binits{W.}},
\oauthor{\bsnm{Mc}, \binits{E.}},
\oauthor{\bsnm{Romero}, \binits{A.M.}}:
Treatment precision of image-guided liver {SBRT} using implanted fiducial
  markers depends on marker-tumour distance.
Physics in Medicine and Biology
(2011)
\end{botherref}
\endbibitem

\bibitem{suzuki_video_1999}
\begin{barticle}
\bauthor{\bsnm{Suzuki}, \binits{K.}},
\bauthor{\bsnm{Nagai}, \binits{K.}},
\bauthor{\bsnm{Yoshida}, \binits{J.}},
\bauthor{\bsnm{Ohmatsu}, \binits{H.}},
\bauthor{\bsnm{Takahashi}, \binits{K.}},
\bauthor{\bsnm{Nishimura}, \binits{M.}},
\bauthor{\bsnm{Nishiwaki}, \binits{Y.}}:
\batitle{Video-assisted thoracoscopic surgery for small indeterminate pulmonary
  nodules}.
\bjtitle{Chest}
\bvolume{115}(\bissue{2}),
\bfpage{563}--\blpage{568}
(\byear{1999})
\end{barticle}
\endbibitem

\bibitem{keating_novel_2016}
\begin{barticle}
\bauthor{\bsnm{Keating}, \binits{J.}},
\bauthor{\bsnm{Singhal}, \binits{S.}}:
\batitle{Novel methods of intraoperative localization and margin assessment of
  pulmonary nodules}.
\bjtitle{Seminars in Thoracic and Cardiovascular Surgery}
\bvolume{28}(\bissue{1}),
\bfpage{127}--\blpage{136}
(\byear{2016})
\end{barticle}
\endbibitem

\bibitem{shieh_bayesian_2017}
\begin{barticle}
\bauthor{\bsnm{Shieh}, \binits{C.C.}},
\bauthor{\bsnm{Caillet}, \binits{V.}},
\bauthor{\bsnm{Dunbar}, \binits{M.}},
\bauthor{\bsnm{Keall}, \binits{P.J.}},
\bauthor{\bsnm{Booth}, \binits{J.T.}},
\bauthor{\bsnm{Hardcastle}, \binits{N.}},
\bauthor{\bsnm{Haddad}, \binits{C.}},
\bauthor{\bsnm{Eade}, \binits{T.}},
\bauthor{\bsnm{Feain}, \binits{I.}}:
\batitle{A {Bayesian} approach for three-dimensional markerless tumor tracking
  using {kV} imaging during lung radiotherapy}.
\bjtitle{Physics in Medicine and Biology}
\bvolume{62}(\bissue{8}),
\bfpage{3065}--\blpage{3080}
(\byear{2017})
\end{barticle}
\endbibitem

\bibitem{zhang_automatic_2020}
\begin{barticle}
\bauthor{\bsnm{Zhang}, \binits{Y.}},
\bauthor{\bsnm{Huang}, \binits{X.}},
\bauthor{\bsnm{Wang}, \binits{J.}},
\bauthor{\bsnm{Sebastian}, \binits{N.}},
\bauthor{\bsnm{Robb}, \binits{R.}},
\bauthor{\bsnm{Webb}, \binits{A.}},
\bauthor{\bsnm{Shilo}, \binits{K.}},
\bauthor{\bsnm{Denicola}, \binits{G.M.}},
\bauthor{\bsnm{Williams}, \binits{T.M.}}:
\batitle{Automatic {Cone} {Beam} {Projection}-based {Liver} {Tumor}
  {Localization} by {Deep} {Learning} and {Biomechanical} {Modeling}}.
\bjtitle{Int. Journal of Radiation Oncology, Biology, Physics}
\bvolume{108}(\bissue{3}),
\bfpage{171}
(\byear{2020})
\end{barticle}
\endbibitem

\bibitem{hirai_real-time_2019}
\begin{barticle}
\bauthor{\bsnm{Hirai}, \binits{R.}},
\bauthor{\bsnm{Sakata}, \binits{Y.}},
\bauthor{\bsnm{Tanizawa}, \binits{A.}},
\bauthor{\bsnm{Mori}, \binits{S.}}:
\batitle{Real-time tumor tracking using fluoroscopic imaging with deep neural
  network analysis}.
\bjtitle{Physica Medica}
\bvolume{59},
\bfpage{22}--\blpage{29}
(\byear{2019})
\end{barticle}
\endbibitem

\bibitem{foote_real-time_2019}
\begin{barticle}
\bauthor{\bsnm{Foote}, \binits{M.D.}},
\bauthor{\bsnm{Zimmerman}, \binits{B.E.}},
\bauthor{\bsnm{Sawant}, \binits{A.}},
\bauthor{\bsnm{Joshi}, \binits{S.C.}}:
\batitle{Real-{Time} {2D}-{3D} {Deformable} {Registration} with {Deep}
  {Learning} and {Application} to {Lung} {Radiotherapy} {Targeting}}.
\bjtitle{Lecture Notes in Computer Science (including subseries Lecture Notes
  in Artificial Intelligence and Lecture Notes in Bioinformatics)}
\bvolume{11492 LNCS},
\bfpage{265}--\blpage{276}
(\byear{2019})
\end{barticle}
\endbibitem

\bibitem{Rouze2016}
\begin{barticle}
\bauthor{\bsnm{Rouze}, \binits{S.}},
\bauthor{\bparticle{de} \bsnm{Latour}, \binits{B.}},
\bauthor{\bsnm{Flecher}, \binits{E.}},
\bauthor{\bsnm{Guihaire}, \binits{J.}},
\bauthor{\bsnm{Castro}, \binits{M.}},
\bauthor{\bsnm{Corre}, \binits{R.}},
\bauthor{\bsnm{Haigron}, \binits{P.}},
\bauthor{\bsnm{Verhoye}, \binits{J.-P.}}:
\batitle{{Small pulmonary nodule localization with cone beam computed
  tomography during video-assisted thoracic surgery: a feasibility study}}.
\bjtitle{Interactive CardioVascular and Thoracic Surgery}
\bvolume{22}(\bissue{6}),
\bfpage{705}--\blpage{711}
(\byear{2016})
\end{barticle}
\endbibitem

\bibitem{lee2022breathing}
\begin{bchapter}
\bauthor{\bsnm{Lee}, \binits{B.C.}},
\bauthor{\bsnm{Sinha}, \binits{A.}},
\bauthor{\bsnm{Varble}, \binits{N.}},
\bauthor{\bsnm{Pritchard}, \binits{W.F.}},
\bauthor{\bsnm{Karanian}, \binits{J.W.}},
\bauthor{\bsnm{Wood}, \binits{B.J.}},
\bauthor{\bsnm{Bydlon}, \binits{T.}}:
\bctitle{Breathing-compensated neural networks for real time c-arm pose
  estimation in lung ct-fluoroscopy registration}.
In: \bbtitle{IEEE International Symposium on Biomedical Imaging (ISBI)},
pp. \bfpage{1}--\blpage{5}
(\byear{2022})
\end{bchapter}
\endbibitem

\bibitem{shen_patient-specific_2019}
\begin{barticle}
\bauthor{\bsnm{Shen}, \binits{L.}},
\bauthor{\bsnm{Zhao}, \binits{W.}},
\bauthor{\bsnm{Xing}, \binits{L.}}:
\batitle{Patient-specific reconstruction of volumetric computed tomography
  images from a single projection view via deep learning}.
\bjtitle{Nature Biomedical Engineering}
\bvolume{3}(\bissue{11}),
\bfpage{880}--\blpage{888}
(\byear{2019})
\end{barticle}
\endbibitem

\bibitem{you_adversarial_2019}
\begin{botherref}
\oauthor{\bsnm{You}, \binits{Z.}},
\oauthor{\bsnm{Ye}, \binits{J.}},
\oauthor{\bsnm{Li}, \binits{K.}},
\oauthor{\bsnm{Xu}, \binits{Z.}},
\oauthor{\bsnm{Wang}, \binits{P.}}:
Adversarial {Noise} {Layer}: {Regularize} {Neural} {Network} by {Adding}
  {Noise}; {Adversarial} {Noise} {Layer}: {Regularize} {Neural} {Network} by
  {Adding} {Noise}.
2019 IEEE International Conference on Image Processing (ICIP)
(2019)
\end{botherref}
\endbibitem

\bibitem{loshchilov_decoupled_2018}
\begin{bchapter}
\bauthor{\bsnm{Loshchilov}, \binits{I.}},
\bauthor{\bsnm{Hutter}, \binits{F.}}:
\bctitle{Decoupled weight decay regularization}.
In: \bbtitle{International Conference on Learning Representations}
(\byear{2018})
\end{bchapter}
\endbibitem

\bibitem{unberath_deepdrr_2018}
\begin{barticle}
\bauthor{\bsnm{Unberath}, \binits{M.}},
\bauthor{\bsnm{Zaech}, \binits{J.N.}},
\bauthor{\bsnm{Lee}, \binits{S.C.}},
\bauthor{\bsnm{Bier}, \binits{B.}},
\bauthor{\bsnm{Fotouhi}, \binits{J.}},
\bauthor{\bsnm{Armand}, \binits{M.}},
\bauthor{\bsnm{Navab}, \binits{N.}}:
\batitle{{DeepDRR} – {A} {Catalyst} for {Machine} {Learning} in
  {Fluoroscopy}-{Guided} {Procedures}}.
\bjtitle{Lecture Notes in Computer Science}
\bvolume{11073 LNCS},
\bfpage{98}--\blpage{106}
(\byear{2018}).
Accessed 2022-02-27
\end{barticle}
\endbibitem

\bibitem{trouve_computing_2005}
\begin{barticle}
\bauthor{\bsnm{Trouve}, \binits{A.}},
\bauthor{\bsnm{Faisal~Beg}, \binits{M.}},
\bauthor{\bsnm{Miller}, \binits{M.I.}},
\bauthor{\bsnm{Younes}, \binits{L.}}:
\batitle{Computing {Large} {Deformation} {Metric} {Mappings} via {Geodesic}
  {Flows} of {Diffeomorphisms}}.
\bjtitle{International Journal of Computer Vision}
\bvolume{61}(\bissue{2}),
\bfpage{139}--\blpage{157}
(\byear{2005})
\end{barticle}
\endbibitem

\bibitem{durrleman_morphometry_2014}
\begin{barticle}
\bauthor{\bsnm{Durrleman}, \binits{S.}},
\bauthor{\bsnm{Prastawa}, \binits{M.}},
\bauthor{\bsnm{Charon}, \binits{N.}},
\bauthor{\bsnm{Korenberg}, \binits{J.R.}},
\bauthor{\bsnm{Joshi}, \binits{S.}},
\bauthor{\bsnm{Gerig}, \binits{G.}},
\bauthor{\bsnm{Trouvé}, \binits{A.}}:
\batitle{Morphometry of anatomical shape complexes with dense deformations and
  sparse parameters}.
\bjtitle{NeuroImage}
\bvolume{101},
\bfpage{35}--\blpage{49}
(\byear{2014})
\end{barticle}
\endbibitem

\bibitem{tobin_domain_2017}
\begin{bchapter}
\bauthor{\bsnm{Tobin}, \binits{J.}},
\bauthor{\bsnm{Fong}, \binits{R.}},
\bauthor{\bsnm{Ray}, \binits{A.}},
\bauthor{\bsnm{Schneider}, \binits{J.}},
\bauthor{\bsnm{Zaremba}, \binits{W.}},
\bauthor{\bsnm{Abbeel}, \binits{P.}}:
\bctitle{Domain {Randomization} for {Transferring} {Deep} {Neural} {Networks}
  from {Simulation} to the {Real} {World}}.
In: \bbtitle{{IEEE}/{RSJ} International Conference on Intelligent Robots and
  Systems ({IROS})},
pp. \bfpage{22}--\blpage{30}
(\byear{2017})
\end{bchapter}
\endbibitem

\bibitem{Banyaga1997}
\begin{bbook}
\bauthor{\bsnm{Banyaga}, \binits{A.}}:
\bbtitle{The Structure of Classical Diffeomorphism Groups}.
\bsertitle{Mathematics and its Applications},
vol. \bseriesno{400}.
\bpublisher{Kluwer Academic}, \blocation{???}
(\byear{1997})
\end{bbook}
\endbibitem

\bibitem{Desai_chest_2020}
\begin{botherref}
\oauthor{\bsnm{Shivang~Desai}, \binits{e.a.}}:
Chest imaging representing a {COVID}-19 positive rural u.s. population.
Scientific Data
\textbf{7}(1)
(2020)
\end{botherref}
\endbibitem

\bibitem{van-de-Kraats_standardized_2005}
\begin{barticle}
\bauthor{\bparticle{van~de} \bsnm{Kraats}, \binits{E.B.}},
\bauthor{\bsnm{Penney}, \binits{G.P.}},
\bauthor{\bsnm{Tomazevic}, \binits{D.}},
\bauthor{\bparticle{van} \bsnm{Walsum}, \binits{T.}},
\bauthor{\bsnm{Niessen}, \binits{W.J.}}:
\batitle{Standardized evaluation methodology for 2-d-3-d registration}.
\bjtitle{IEEE Transactions on Medical Imaging}
\bvolume{24}(\bissue{9}),
\bfpage{1177}--\blpage{1189}
(\byear{2005})
\end{barticle}
\endbibitem

\bibitem{hugo_longitudinal_2017}
\begin{barticle}
\bauthor{\bsnm{Hugo}, \binits{G.D.}},
\bauthor{\bsnm{Weiss}, \binits{E.}},
\bauthor{\bsnm{Sleeman}, \binits{W.C.}},
\bauthor{\bsnm{Balik}, \binits{S.}},
\bauthor{\bsnm{Keall}, \binits{P.J.}},
\bauthor{\bsnm{Lu}, \binits{J.}},
\bauthor{\bsnm{Williamson}, \binits{J.F.}}:
\batitle{A longitudinal four-dimensional computed tomography and cone beam
  computed tomography dataset for image-guided radiation therapy research in
  lung cancer}.
\bjtitle{Medical physics}
\bvolume{44}(\bissue{2}),
\bfpage{762}
(\byear{2017})
\end{barticle}
\endbibitem

\end{thebibliography}

\end{document}